\documentclass[10pt]{article}
\usepackage[margin=1in]{geometry}
\usepackage{graphicx, amsthm, enumerate, enumitem, amsmath, amssymb, siunitx, microtype} 
\usepackage[title,titletoc]{appendix} 
\usepackage[section]{placeins}
\newcommand{\sign}{\text{sgn}}
\usepackage[
  backend=biber,
  style=ieee,
  minnames=1,
  maxcitenames=2, maxbibnames=6
]{biblatex}
\bibliography{ref.bib}

\title{A Safe and Stable Controller for Fuel Cell Systems Using Adaptation and Reference Governors}
\author{Mychal Amoafo, Ilya Kolmanovsky, and Anuradha Annaswamy }
\date{April 20, 2026}
\begin{document}
\maketitle

\begin{abstract}                
This paper proposes a control architecture integrating adaptation with Lyapunov-based Reference Governors (LRGs) to ensure state constraint satisfaction for first-order systems with parametric uncertainties. Adaptation combined with LRGs guarantees stability, ensures good control performance, and remains safe even with parametric uncertainties. Simulations of the fuel cell temperature regulation problem demonstrate that the proposed control architecture successfully meets all control and safety objectives, whereas the standard adaptation fails to achieve the latter. 
\end{abstract}


\section{Introduction}

Fuel cell systems are becoming an integral part of energy systems like microgrids where they improve reliability and resilience. In such applications, thermal management is crucial to maximizing efficiency, maintaining consistent power output, and extending lifespan. An ideal thermal management system ensures that a series of temperature set points are accurately tracked, despite changes in the fuel cell power. Large changes in the fuel cell stack temperature caused by variations in power can cause hot spots within the fuel cell stack, contributing to rapid material degradation, reductions in efficiency and operational lifespan.

In this larger context, it becomes clear that two system properties are simultaneously important for a thermal management system in fuel cells: regulation and safety. First, regulation performance has to be ensured, around a range of set points that may be dictated by the microgrid. Second, safe regulation has to be ensured around this set point, in that any excursions around the desired set point have to be maintained within the prescribed range. Finally, both regulation and safety have to be achieved in the presence of disturbances and modeling uncertainties. 

There is extensive literature addressing the design of temperature controllers for fuel cell systems. In \cite{kurz_temp_air_pi}, the authors design a proportional integral (PI) controller to regulate the temperature and air flow rate of a portable, high temperature proton-exchange membrane fuel cells (PEMFC). In \cite{pemfc_phd_fr}, the author designed and experimentally validated a state feedback controller from a control-oriented state-space model of the fuel cell temperature dynamics around a nominal operating condition. 
In \cite{mrac_pemfc}, the authors developed a model reference adaptive controller (MRAC) to account for uncertainties caused by changing operation conditions. These uncertainties included variations in the electric load and changing ambient temperature. In \cite{netl_mrac_tucker}, the authors proposed a MRAC to regulate the turbine speed and cathode inlet mass flow rate in a solid-oxide fuel cell gas turbine hybrid system, which is shown to lead to improvement in performance. In all of the above papers, the authors do not consider constraint enforcement in the control design. 

The approach in \cite{ossareh_rg_fuelcells} addresses constraints due to pressure in the context of a water electrolysis system, and utilizes RGs in conjunction with a state feedback controller to realize their objectives. No modeling uncertainties are however addressed in this paper. In \cite{L1_adaptive_rg}, the authors propose an integrated adaptive control and reference governor  where an $\mathcal{L}_1$ adaptive controller is used to compensate for uncertainties and a reference governor is used to enforce constraints. The combination of adaptive control and reference governors in \cite{L1_adaptive_rg} used a maximum output admissible set precomputed offline; in contrast, our proposed approach allows for online computation of  a constrained admissible set as a sub-level set of a Lyapunov function. 

In this paper, we propose a safe and stable controller that combines an adaptive controller with a Lyapunov-based reference governor (LRG) to achieve set point regulation  and  satisfaction of constraints due to temperature limits in the presence of parametric uncertainties. 

The rest of this article is organized as follows. In Section 2, we describe the problem statement. Section 3 includes an overview of a range of underlying tools related to the adaptive PI and LRG approaches. Section 4 describes the safe-and-stable adaptive controller for Problem 1. Section 5 includes numerical study of the safe-and-stable adaptive controller for Problem 1. Proofs of Theorem 1 and all lemmas can be found in the Appendixes.

\section{The fuel cell model}
In this section, we present a reduced-order linear model that determines the relationship between the mass flow rate of coolant and temperature of the fuel cell stack. While CFD models can be employed to understand the relationship between the mass flow rate of the coolant and the temperature of the fuel cell stack, they are too complex and not tractable for designing a control system. The following assumptions are made simplify this relationship and produce a reduced-order linear model: 
\begin{itemize}
    \item Uniform temperature distribution across the fuel cell stack. 
    \item Negligible heat loss due to natural convection and radiation. 
    \item The heat exchange due to the gaseous species is neglected. 
\end{itemize}

Using these assumptions, thermal balance, constitutive relations, Faraday's law of electrolysis, Ohm's Law, we can derive the following equations
\begin{equation}\label{eq: pemfc_thermal_balance}
     \frac{dT_{st}(t)}{dt} = \frac{1}{m_{st}C_{p_{st}}}\Big(\dot Q_{gen} - P_{elec} - \dot Q_{cool}\Big)         
\end{equation}
\begin{equation}\label{eq: pemfc_total_power}
    \dot Q_{gen} =  \dot m_{H_2}\Delta H 
\end{equation}
\begin{equation}\label{eq: pemfc_current}
        I_{cell} = \frac{2F}{\eta_{I}N_{cell}MM_{H_2}} \dot m_{H_2} 
\end{equation}
\begin{equation}\label{eq: pemfc_electrical_power}
         P_{elec} = N_{cell}V_{cell}I_{cell} 
\end{equation}
\begin{equation}\label{eq: pemfc_voltage}
           V_{cell} = E_0 - V_{act} - V_{ohm} - V_{trans} 
\end{equation}
\begin{equation}\label{eq: pemfc_coolant_power}
    \dot Q_{cool} = w_c C_{p_c}\Big(T_{st}  - T_{in}\Big)
\end{equation}
All quantities in \eqref{eq: pemfc_thermal_balance}-\eqref{eq: pemfc_coolant_power} are defined in Appendix A and are used to derive the following bilinear model
\begin{equation}\label{eq: pemfc_bilinear}
    \frac{dT_{st}(t)}{dt} = A_0(I_{cell}) + A_1(I_{cell})T_{st}(t) - w_{c} B_0 \left(T_{st}(t) - T_{in}\right)
\end{equation}
where $T_{st}$ denotes the stack temperature, and $w_c$ denotes the coolant mass flow rate. We note that $A_0(I_{cell}) + A_1(I_{cell})T_{st}(t)$ can be viewed as the difference between $\dot Q_{gen}$ and $P_{elec}$ at a specific operating point defined by load current and stack temperature, $I_{cell}^\circ$ and $T_{st}^\circ$, respectively. This is used to determine a nominal control input $w_{c}^\circ$ required to remove the excess heat generated by the fuel cell stack, which can be defined as
\begin{equation}\label{eq: nominal_coolant_flowrate}
    w_c^\circ = \frac{A_0(I_{cell}^\circ) + A_1(I_{cell}^\circ)T_{st}^\circ }{B_0(T_{st}^\circ - T_{in})}
\end{equation}
Considering the temperature dynamics of the fuel cell operating near $(T_{st}^\circ, w_{c}^\circ)$, we linearize the bilinear model in \eqref{eq: pemfc_bilinear} as 
\begin{equation}\label{eq: linearized_pemfc}
    \dot x(t)  = \frac{1}{J}\Big(-B x(t) + u(t)\Big)
\end{equation}
where $x(t) = T_{st} - T_{st}^\circ$  and $u(t) = w_c - w_c^\circ$, 
\begin{equation}\label{eq: pemfc_parameters}
    J = -\frac{1}{B_0(T_{st}^\circ - T_{in}^\circ)}, \quad B = -\frac{B_0w_c^\circ - A_1(I_{cell}^\circ)}{B_0(T_{st}^\circ - T_{in}^\circ)}.
\end{equation}
The specific problem we consider in this paper is the design of $u(t)$ in \eqref{eq: linearized_pemfc} to maintain $x(t)$ within the prescribed range when $B$ and $J$ are unknown due to parametric uncertainties. Sources of uncertainties that can, to an extent, be accounted for by the parametric uncertainty are part-to-part variability, material degradation, changes in operating conditions, and model mismatch.  

The definition of $J$ in \eqref{eq: pemfc_parameters} implies that $J<0$ for all values of $B_0$. Since $B_0>0$ and $T_{st}^\circ > T_{in}$ which follows since $T_{in}$ is sufficiently cooled. In all that follows, $J<0$ for all operating conditions.   

\subsection{State constraints}

An important consideration in the thermal management of fuel cells is the need to ensure that limits on the temperature of the fuel cell stack are met. That is, in addition to designing $u(t)$ so that $x(t)$ follows a specific temperature setpoint, we also need to make sure that $|x(t)| \leq \bar x$ where $\bar x$ represents a maximum allowable deviation away from $T_{st}^\circ$. 

\subsection{Statement of the problem}
The problem addressed in this paper is the design of a control law for $u(t)$ in \eqref{eq: linearized_pemfc} where $B, J$ are unknown, so that a \textit{safety objective} and \textit{control objective} are met. These objectives are defined as follows: The \textit{safety objective} is to ensure that $x(t) \in \mathcal{C}_{\bar x}$ where
\begin{equation}\label{eq: admissible_set}
    \mathcal{C}_{\bar x} = \{x : |x| \leq \bar x \quad \forall t\geq t_0\}.
\end{equation}
The \textit{control objective} is set point regulation, that is, for $x(t)$ to follow a constant reference command $x_d$ asymptotically. The solution to this problem will be addressed in Section 4.  

\section{Preliminaries}
The safe and stable adaptive controller that we propose in this paper is built using two well known tools: adaptive control and reference governors.

\subsection{Adaptive PI control}
Adaptive control is a mature, advanced control methodology applicable to the control of dynamic systems with parametric uncertainties. The controller uses a parameter estimation algorithm that is designed such that the fundamental properties of stability, convergence, robustness, and learning are ensured and established with analytical guarantees. In what follows, we develop an adaptive PI controller where the PI gains are adjusted using a prescribed law which results in a closed-loop system with desirable tracking properties (\cite{adaptivePI,stableAdaptiveSystems}). 

The problem we will consider is a design of a PI controller of the plant in \eqref{eq: linearized_pemfc} when the parameters $B$ and $J$ are unknown. Towards this end, we will first develop a control solution when $B$ and $J$ are at their nominal values $B_{nom}$ and $J_{nom}$, respectively. This will be followed by an adaptive controller design that consists of  estimates of $B$ and $J$ so that the closed-loop system has bounded solutions and $x(t)$ tracks a desired signal asymptotically. 

\begin{figure}[htbp]
    \centering
    \includegraphics[width=0.75\linewidth]{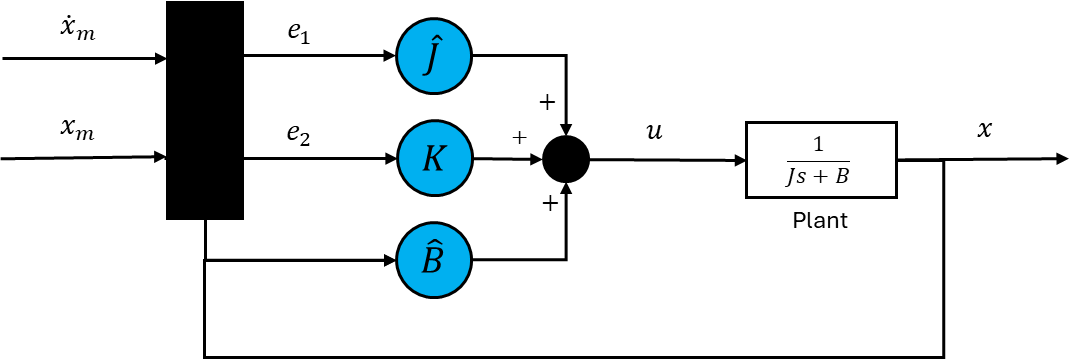}
    \caption{Adaptive PI control}
    \label{fig:adaptive pi}
\end{figure}

When $B_{nom}$ and $J_{nom}$ are known, the plant model takes the form
\begin{equation}\label{eq: ref_plant}
    \dot x_m(t) = \frac{1}{J_{nom}}\bigg(-B_{nom}x_m(t) + u(t)\bigg)
\end{equation}
We choose a PI controller of the form (see Figure \ref{fig:adaptive pi})
\begin{equation}\label{eq: pi_control_ref}
    u(t) = B_{nom}x_m (t) + J_{nom}e_{1m}(t) + Ke_{2m}(t)
\end{equation}
where 
\begin{equation}\label{eq: ref_mod_errors}
    \begin{split}
        e_m(t) = x_d(t) - x_m(t), \quad  e_{1m}(t) = \dot x_d(t) + \lambda e_m(t)
    \end{split}
\end{equation}
\begin{equation}\label{eq: error_sum_model}
    e_{2m}(t) = e_m(t) +\lambda \int_{0}^{t}e_m(\tau) d\tau,
\end{equation}
$K < 0$ and  $\lambda> 0$. It follows from \eqref{eq: ref_plant}-\eqref{eq: error_sum_model} that  $e_{2m}(t)$ in \eqref{eq: error_sum_model} satisfies the differential equation
\begin{equation}\label{eq: ref_model_closedloop}
    \dot e_{2m}(t) = -\frac{K}{J_{nom}}e_{2m}(t)
\end{equation}
It is easy to see from \eqref{eq: ref_model_closedloop} that $e_{2m}(t)$ and $\dot e_{2m}(t)$ tend to zero and then from \eqref{eq: error_sum_model} that, $x_m(t)$ tracks $x_d(t)$ asymptotically.  We also note that \eqref{eq: ref_plant}-\eqref{eq: error_sum_model}, under the assumed conditions on the gains, represents a stable closed-loop system. We will refer to this closed-loop system as a reference model, as it provides a satisfactory solution when the plant parameters are at their known nominal values. 

We now return to the control of the plant in \eqref{eq: linearized_pemfc} where the challenge is the design of a control law similar to \eqref{eq: pi_control_ref} when the plant parameters may not be at their nominal values and could have changed due to a variety of reasons including the aging of components or variations in the operating conditions. We choose 
\begin{equation}\label{eq: adaptive_control_law}
    u(t) =  \widehat B(t)x(t) + \widehat J(t)e_1 (t) + Ke_2(t)
\end{equation}
where  
\begin{equation}\label{eq: adaptive_errors}
    \begin{split}
        e_{1}(t) = \dot x_{m}(t) + \lambda e(t)\\  e_2(t) = e(t) + \lambda \int_{0}^{t} e(\tau) d\tau \\ e(t) = x_m(t) - x(t)
    \end{split}
\end{equation}
leads to a stable closed-loop system with suitable adaptive laws for adjusting $\widehat B$ and $\widehat J$ which are estimates of the unknown parameters $B$ and $J$, respectively. Such a stability property follows as shown below. The above controller leads to the following closed-loop system (\cite{adaptivePI}):
\begin{equation}\label{adaptive_pi_closedloop}
    \dot e_2(t) = -\frac{K}{J}e_2(t) + \frac{1}{J}\left(-\widetilde Je_1(t) - \widetilde Bx(t)\right) ,
\end{equation}
where
\begin{equation*} \label{eq: parameter_errors}
    \widetilde J =  \widehat J -  J,\quad \widetilde B = \widehat B - B.
\end{equation*}
denote the parameter estimation errors. We now propose the following adaptive laws: 
\begin{equation}\label{eq: adaptive_laws}
\begin{split}
    \dot{\widehat{J}} = \gamma_1e_1e_2, \quad 
   \dot{\widehat{B}} = \gamma_2xe_2
\end{split}
\end{equation}
where $\gamma_1, \gamma_2 <0$. The stability of the complete closed system can be determined through a Lyapunov function candidate
\begin{equation}\label{eq: lyapunov_function_candidate}
    V = \frac{1}{2}\left(e_2^2 + \frac{1}{J}\left(\frac{\widetilde J^2}{\gamma_1} + \frac{\widetilde B^2}{\gamma_2} \right)\right)
\end{equation}as it leads to 
\begin{equation}\label{eq: lyapunov_function_derivative}
    \dot V = -\frac{K}{J}e_2^2 \leq 0. 
\end{equation}
Boundedness of $e_2, e_1$, and $e$ follows. Standard arguments using Barbalat's lemma and continuity ensures that $x(t)$ tracks $x_m(t)$ asymptotically.

\subsection{Lyapunov-based reference governors}

Reference governors (RGs) are add-on safety filters that enforce pointwise-in-time state and input constraints by modifying reference commands to a stable closed-loop system. Operating between the reference command and the input to the closed-loop system, RGs allow nominal control design to be separated from the task of constraint satisfaction. Over the years, several approaches have been proposed for designing RGs, which were initially developed for linear systems (\cite{GILBERT20022063}). In what follows, we will focus on LRGs which are applicable for nonlinear systems. 

Unlike standard RG approaches which utilize \textit{maximal constraint admissible sets}, LRGs translate pointwise-in-time state constraints into an upper bound on a Lyapunov function parameterized by a reference so that the corresponding sublevel set describes a domain of attraction around an equilibrium point of a stable closed-loop system (\cite{nonlinearRGConference}).  LRGs have been explored for a variety of applications (\cite{refgov_survey}). We briefly describe the LRG approach below. 

Consider the following nonlinear system
\begin{equation} \label{eq: nonlineardynamics}
    \dot x (t)=  f(x(t),v(t)) 
\end{equation}
where $x(t) \in \mathbb{R}^n$ is the state and $v(t) \in \mathbb{R}^m$ is the input. It is assumed that \eqref{eq: nonlineardynamics} represents a closed-loop system which is stable and has good tracking performance (\cite{nonlinearRGConference}).  The constraints on \eqref{eq: nonlineardynamics} are specified as 
\begin{equation}\label{eq: nonlinear_constraints}
    x(t) \in \mathcal{S} \hspace{5mm} \forall t\geq 0
\end{equation}
where $\mathcal{S} $ is the admissible region prescribed by pointwise-in-time state and control. We define a Lyapunov function $V(x,r)$ where $r$ is an external command. More specifically,  $V(x,r)$ is chosen to correspond to a stable equilibrium $x_e(r)$ of \eqref{eq: nonlineardynamics} and so that $V(x_e(r),r)=0 $ and $V(x,r) >0$ for all $x\neq x_e(r)$.  Next, we define $\Gamma(r)$ which is chosen such that all trajectories of \eqref{eq: nonlineardynamics} satisfy the constraint \eqref{eq: nonlinear_constraints}. To accomplish this, we define $\Gamma(r)$ as 
\begin{equation}\label{eq: largest_V}
    \Gamma (r) = \min_x V(x,r) \text{ such that } x \not \in S
\end{equation}
With such a definition in \eqref{eq: largest_V}, it follows from the fact $V(x,r)$ is a Lyapunov function that 
\[V(x(t_0, r) \leq \Gamma (r) \implies V(x(t), r)) \leq \Gamma(r) \quad \forall t\geq t_0\]
The LRG can now be completely specified by connecting the choice of $v$ to $r$ through a parameterized representation as follows (\cite{nonlinearRGConference}): 
\begin{equation*}
    v(t) = v(t-1) + \kappa(t) \Big(r(t) - v(t-1)\Big)
\end{equation*}
It can be seen that when $\kappa(t)=1$, $v(t)=r(t)$, which implies that the control input is identical to the reference input, and when $\kappa(t)=0$, the reference is entirely disconnected, and $v(t)=v(t-1)$. For any value of $\kappa (t) \in (0,1)$, $v(t)$ takes on an interpolated value given by $v(t)=\kappa (t)r(t)+ (1-\kappa(t))  v(t-1)$. 

The LRG determines $\kappa(t)$ through the solution of the following optimization problem: 
\begin{subequations}\label{eq: optimization_problem}
\label{eq: optimization_group}
\begin{align}
    \max_{\kappa \in [0,1]}\quad & \kappa \label{eq: a}\\ 
    \textrm{s.t.} \quad & v(t) = v(t-1) + \kappa (r(t) - v(t-1))            \label{eq: b}\\
     & V(x(t), v(t)) - \Gamma(v(t)) \leq 0 \label{eq: c} 
\end{align}
\end{subequations}
Assuming that $v(t_0)$ exists such that the initial conditions satisfy the inequality $V(x(t_0), v(t_0)) \leq  \Gamma(v(t_0))$ and there is no model mismatch \eqref{eq: optimization_problem} remains feasible for all $t\geq t_0$ and the constraints are guaranteed to be satisfied. 

The complexity of designing an LRG is entirely related to the choice of a suitable $\Gamma(r)$, a suitable $V(x,r)$, and ensuring that a feasible $\kappa$ exists that satisfies the constraints in \eqref{eq: optimization_problem_xd}.  In general, the approach outlined in this section is not feasible for all nonlinear systems because $V(x,r)$ and $\Gamma(r)$ can be difficult to determine. To overcome this issue, a few strategies can be found in \cite{GILBERT20022063}. 

In what follows, we consider a special case of the LRG when $r(t) = r_0$, a constant. 
\newline \textbf{Proposition 1.} There exists a set $R=\{r: |r_0| \leq r_{max}\}$ for which $\kappa (t) \to  1$ and $v(t) \to r_0$ as $t \to \infty $. 
\newline \textbf{Proof.} The proof of Proposition 3.1 follows from Theorems 1 and 2 in \cite{GILBERT20022063}.

\section{A safe and stable adaptive controller}

We now present the main result of this paper which is a safe and stable adaptive control design. 

\subsection{A reference governor for a reference model}
\begin{figure}[htbp]
    \centering
    \includegraphics[width=0.75\linewidth]{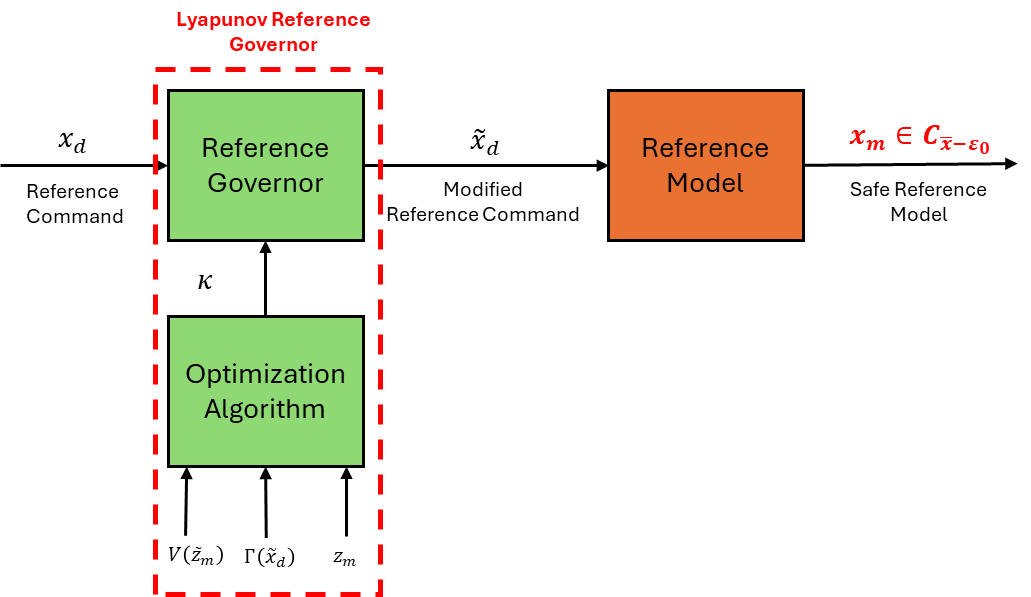}
    \caption{Design of a safe reference model}
    \label{fig: safe_reference_model}
\end{figure}

We first design a safe and stable reference model 
by constructing a LRG shape the reference input. The safety of the reference model is quantified as the requirement that $x_m \in \mathcal{C}_{\bar x}$ where $\mathcal{C}_{\bar x}$ is defined in \eqref{eq: admissible_set}.

We consider the reference model defined by \eqref{eq: ref_plant}-\eqref{eq: pi_control_ref} which can be rewritten, after algebraic manipulations, as 
\begin{equation}\label{eq: ref_model_statespace}
    \begin{bmatrix}
        \dot x_m \\ \dot e_{Im}
    \end{bmatrix} = \begin{bmatrix}
       - \frac{K + J_{nom}\lambda}{J_{nom}} & \frac{K\lambda}{J_{nom}} \\
        -1 & 0
    \end{bmatrix}\begin{bmatrix}
        x_m \\ e_{Im}
    \end{bmatrix} + 
    \begin{bmatrix}
        \frac{K + J_{nom}\lambda}{J_{nom}} & 1   \\ 1 & 0 
    \end{bmatrix} \begin{bmatrix}
        x_d \\ \dot x_d
    \end{bmatrix}
\end{equation}
where $\dot e_{Im} = e_m $. For the case when $x_d$ is a constant \eqref{eq: ref_model_statespace} can be rewritten as 
\begin{equation}\label{eq: ref_model_statespace_2}
    \begin{bmatrix}
        \dot x_m \\ \dot e_{Im}
    \end{bmatrix} = \begin{bmatrix}
       - \frac{K + J_{nom}\lambda}{J_{nom}} & \frac{K\lambda}{J_{nom}} \\
        -1 & 0
    \end{bmatrix}\begin{bmatrix}
        x_m \\ e_{Im}
    \end{bmatrix} + 
    \begin{bmatrix}
        \frac{K + J_{nom}\lambda}{J_{nom}}   \\ 1 
    \end{bmatrix} x_d 
\end{equation}
and simplified as:
\begin{equation*}
    \dot z_m = A_m z_m + b_mx_d
\end{equation*}
where $A_m$ is Hurwitz,  $z_m = [x_m, e_{Im}]^T$ is the state of the reference model, and $x_d$ is a constant. Lemma A.1 implies that $x_m(t)$ converges to $x_d$ as $t \to \infty$ for any initial conditions.

We now modify the external command $x_d$ to $\tilde x_d$  using an LRG described in detail below so as to ensure safety of the the reference model. This leads to the following dynamics: 
\begin{equation}
    \dot z_m = A_m  z_m + b_m \tilde x_d \label{eq: refmodel}
\end{equation}
Next, we choose a Lyapunov function
\begin{equation}\label{eq: V_safety}
    V(\tilde z_m) = \tilde z_m^TP\tilde z_m
\end{equation}
where $\tilde z_m = z_m - \bar z_m$, $\bar z_m = -A_m^{-1}b_m \tilde x_d$, and $P$ is the solution to the Lyapunov equation $A_m^TP + PA_m = -Q$ with $Q = Q^T>0$. To ensure safety of the reference model, we have to satisfy $x_m \in \mathcal{C}_{\bar x}$. Towards this end, we introduce a buffer $\varepsilon_0$ and require that $x_m \in \mathcal{C}_{\bar x - \varepsilon_0}$ with $\varepsilon_0$ such that  $0 < \varepsilon_0 < \bar x$ so $\mathcal{C}_{\bar x - \varepsilon_0}$ is not empty.

We solve the following modified optimization problems to guarantee $x_m \in \mathcal{C}_{\bar x - \varepsilon_0} $ (Figure \ref{fig: safe_reference_model}). First we define
\begin{equation}\label{eq: largest_V_x}
    \Gamma_{\varepsilon_0}(\tilde x_d) = \min _{z_m} V(\tilde z_m) \text{ such that } x_m \not \in \mathcal{C}_{\bar x - \varepsilon_0}.
\end{equation}
We note that  \eqref{eq: largest_V_x} is explicitly solvable using KKT and has a closed-form solution. The following optimization algorithm determines a $\kappa(t)$ such that $V(\tilde z_m) - \Gamma_{\varepsilon_0}(\tilde x_d(t)) \leq 0 \implies x_m \in \mathcal{C}_{\bar x- \varepsilon_0}$: 
\begin{subequations}\label{eq: optimization_problem_xd}
\begin{align}
    \max_{\kappa \in [0,1]} \quad & \kappa\\ 
    \textrm{s.t.} \quad & \tilde x_d(t) = \tilde x_d(t-1) + \kappa \Big(x_d(t) - \tilde x_d(t-1)\Big)            \label{eq: optimization_problem_xd_b}\\
     & V(\tilde z_m) - \Gamma_{\varepsilon_0}(\tilde x_d(t)) \leq 0 \label{eq: optimization_problem_xd_c}
\end{align}
\end{subequations}

This ensures that $\tilde x_d$ remains close to $x_d$  so that $x_m$ can track $x_d$ as closely as possible. Therefore, the reference model in \eqref{eq: ref_model_statespace} meets the control objective while the LRG guarantees that \eqref{eq: ref_model_statespace} meets the safety objective. 

\subsection{A safe and stable adaptive controller}
\begin{figure}[htbp]
    \centering
    \includegraphics[width=1.0\linewidth]{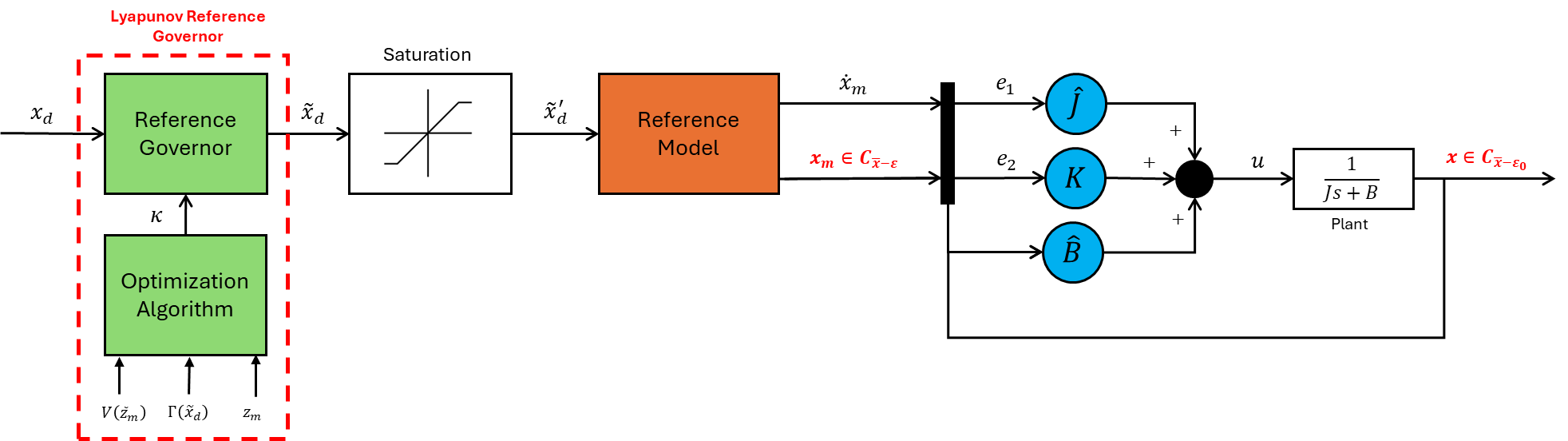}
    \caption{A safe and stable adaptive controller}
    \label{fig:safe_stable_adaptive_controller}
\end{figure}
We now return to the problem posed in Section 2.2, which is the design of $u(t)$ in \eqref{eq: linearized_pemfc} so that the safety and control objectives defined in Section 2.2 are met by the closed-loop adaptive system when $B$ and $J$ are unknown. 

To accomplish the safety objective, we add an extra layer of buffer to the safe set, choosing it as $\mathcal{C}_{\bar x - \varepsilon}$ where 
\begin{equation}\label{time_varying_buffer}
    \varepsilon  = \varepsilon_0 + k_{\varepsilon}e^2
\end{equation}
where $e$ is defined by \eqref{eq: adaptive_errors}. The rationale for this is that the closed-loop adaptive system may not be safe even if the reference model is safe due to the adaptation errors in the state. Therefore, increasing the buffer defined by \eqref{time_varying_buffer} ensures that the buffer is larger to accommodate for the transient errors. That is, when $e$ is large, the buffer increases to ensure safety. As the adaptive controller converges, $e$ becomes smaller, which in turn causes the buffer to shrink, with $\varepsilon \rightarrow \varepsilon_0$. What remains to be shown is that if $x_{m} \in \mathcal{C}_{\bar x - \varepsilon}$ then $x \in \mathcal{C}_{\bar x -\varepsilon^{'}_0}$, where $\varepsilon^{'}_0 = \varepsilon_0 - \delta$ for an arbitrarily small value of $\delta$. 

Before we proceed with the main result, 
we introduce a modification to the constraints applied to $\kappa$ in \eqref{eq: optimization_problem_xd} by allowing $\kappa \in [-\Delta,1]$ rather than limiting it to $[0,1]$. This is needed to allow $\tilde x_d$ to be non-monotonic during transients 
as the constraints $x \in \mathcal{C}_{\bar x - \varepsilon}$ can become less or more strict depending on $e$. 

The change to the constraints on $\kappa$, however, introduces a possibility of $\tilde x_d$ determined by \eqref{eq: optimization_problem_xd_b} becoming unbounded. 
In order to circumvent this, 
we introduce a saturation block as follows (see Figure \ref{fig:safe_stable_adaptive_controller} for a schematic) (\cite{adaptivePI}):
\begin{equation}\label{eq: saturation_filter}
    \tilde x^{'}_{d}(t) = \begin{cases}
        \tilde x_d (t) & \text{if} \quad |\tilde x_d(t)| \leq \bar x_d \\ \bar x_d \sign (\tilde x_d(t)) & \text{if} \quad |\tilde x_d(t)| > \bar x_d
    \end{cases}
\end{equation}
where $|\bar x_d| \in \mathcal{C}_{\bar x - \varepsilon_0} $. We now summarize the overall safe and stable adaptive controller for the plant in \eqref{eq: linearized_pemfc}. We choose $u(t)$ as in \eqref{eq: adaptive_control_law} with $e_1$ and $e_2$ defined as in
\eqref{eq: adaptive_errors}, $\dot x_m$ is determined using \eqref{eq: ref_model_statespace} as follows:
\begin{equation}\label{eq: ref_model_trajectories}
\dot x_m = -\frac{K+J_{nom}\lambda}{J_{nom}} x_m +\frac{K\lambda}{J_{nom}}e_{Im} + \frac{K+J_{nom}\lambda}{J_{nom}} \tilde x^{'}_{d},
\end{equation}
the parameter estimates are given in \eqref{eq: adaptive_laws}  and $\tilde x_{d}$ is determined by solving the following optimization problem: 
\begin{subequations}\label{eq: optimization_problem_xd_nm}
\begin{align}
    \max_{\kappa \in [-\Delta,1]} \quad & \kappa \label{eq: optimization_problem_xd_nm_a}\\ 
    \textrm{s.t.} \quad & \tilde x_d (t)= \tilde x_d(t-1) + \kappa \Big(x_d(t) - \tilde x_d(t-1)\Big)            \label{eq: optimization_problem_xd_nm_b}\\
     & V(\tilde z_m) - \Gamma_{\varepsilon}(\tilde x_d(t)) \leq 0. \label{eq: optimization_problem_xd_nm_c}
\end{align}
\end{subequations}
where $\Delta >0$ and $\varepsilon$ is defined in \eqref{time_varying_buffer}. The following proposition is needed to quantify the impact of the two changes made above.
\newline \textbf{Proposition 2.} The dynamic system specified by \eqref{eq: optimization_problem_xd_nm_a}, \eqref{eq: optimization_problem_xd_nm_b}, and \eqref{eq: saturation_filter} has the following properties: (i) $\tilde x^{'}_d(t)$ is bounded and (ii) if $\tilde x_d(t)$ tends to $x_d$ asymptotically, then $\tilde x^{'}_d(t)$ tends to $x_d$ asymptotically.
\newline \textbf{Proof.} The proof of Proposition 2 follows in a straightforward manner due to the saturation function and its continuity. 
We now state the main result in Theorem 1 whose proof can be found in Appendix C, which utilizes a few lemmas stated and proved in Appendix B.
\newline \textbf{Theorem 1.} If $\tilde x_d(t)$ tends to $x_d$ asymptotically then the closed-loop system defined by \eqref{eq: linearized_pemfc}, \eqref{eq: adaptive_control_law}, \eqref{eq: adaptive_errors}, \eqref{eq: adaptive_laws}, \eqref{eq: ref_model_statespace}, \eqref{eq: saturation_filter}, \eqref{eq: ref_model_trajectories}, and \eqref{eq: optimization_problem_xd_nm} (i) has globally bounded solutions, (ii) meets the control objective, and (iii)  $x(t) \in \mathcal{C}_{\bar x - \varepsilon^{'}_{0}}$ for all $t \geq 0$ and $\frac{1}{4k_\varepsilon}<\delta <\varepsilon_0$.

The sufficient condition in Theorem 1 
is a strong assumption but is not unreasonable  as the error $e(t)$ is expected to converge to zero as $t \to \infty$ so that the reference governor starts functioning as the conventional reference governor for the case of time-independent constraints.  Future work will focus on relaxing this assumption. 

\section{Simulation}
\begin{figure}[htbp]
    \centering
    \includegraphics[width=1\linewidth]{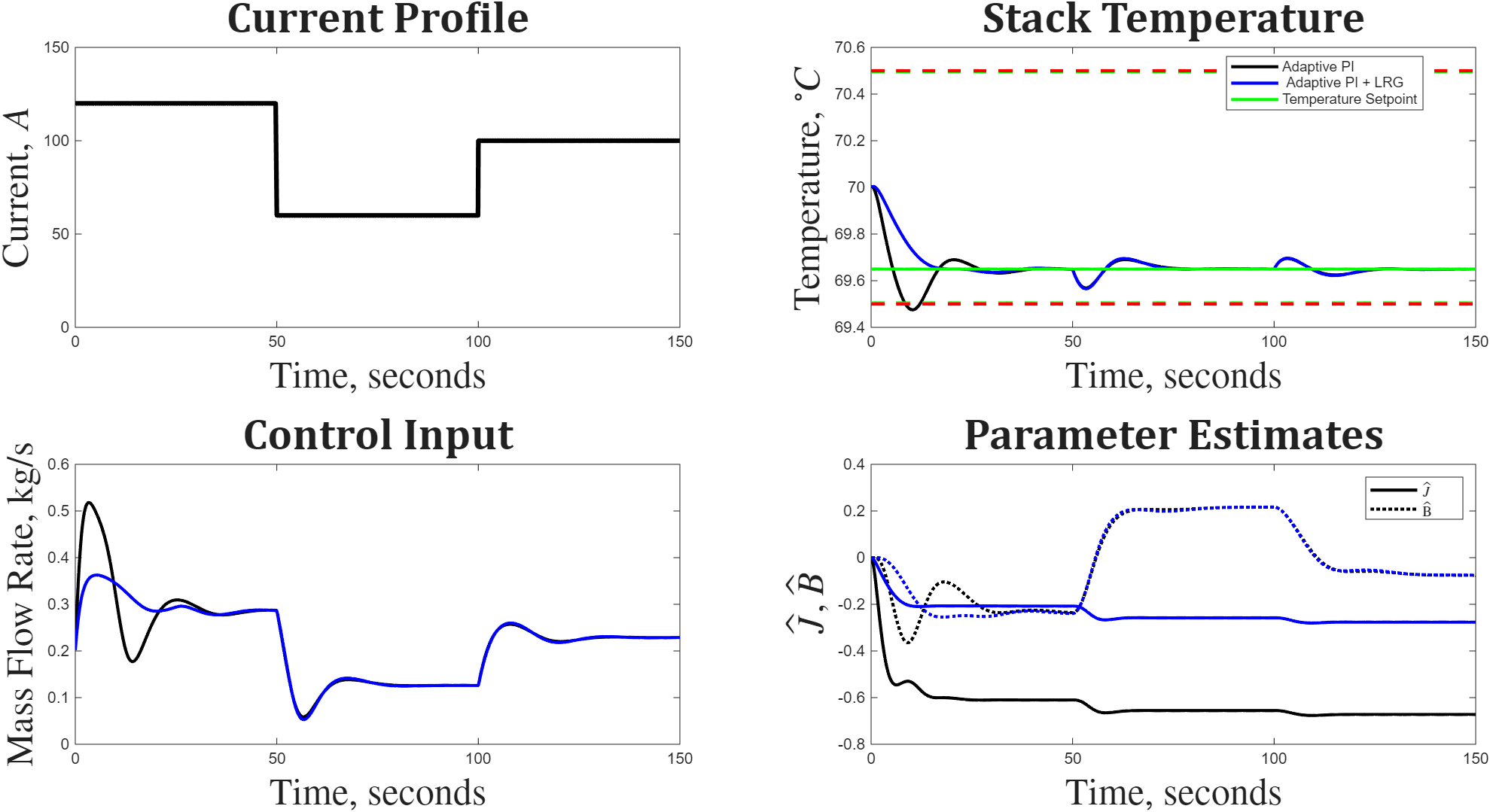}
    \caption{Comparison of the fuel cell stack temperature for the safe and stable adaptive controller and the adaptive controller.}
    \label{fig:temperature_profile}
\end{figure}

\begin{figure}[htbp]
    \centering
    \includegraphics[width=1\linewidth]{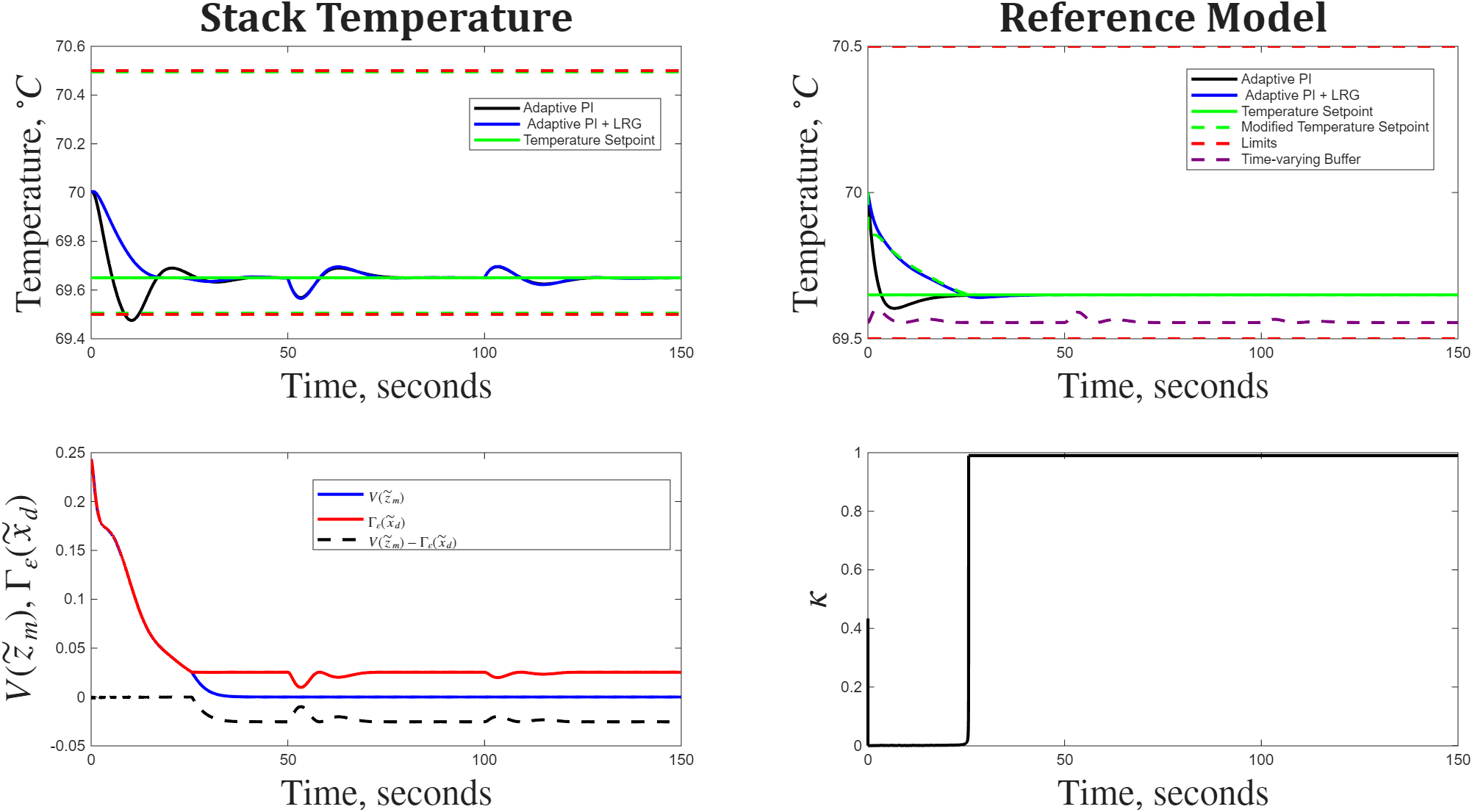}
    \caption{Comparison of the reference trajectories for the safe and stable adaptive controller and the adaptive controller.}
    \label{fig:temperature_reference_model}
\end{figure}

To demonstrate the efficacy of our proposed approach, we apply the safe and stable adaptive controller to a temperature control subsystem in a 5 kW PEMFC system. The parameters of the fuel cell system is provided in Appendix A. 
An operating point $T_{st}^\circ = \SI{343.15}{\kelvin}$, $w_{c}^\circ = \SI{0.20}{\kilo\gram\per\second}$, and $I_{cell}^\circ = \SI{100}{\ampere}$,  a constant temperature setpoint $x_d = \SI{-0.35}{\celsius}$, and a safety constraint of $\bar x = \SI{0.5}{\celsius}$ are chosen. This leads to nominal values $J= -2.79$ , $B = -0.07$, $K=-1.0 $. Parametric uncertainties 11\% and 22\% were introduced in $J$ and $B$ respectively. The PI control parameters   $K=-1.0 $ and $\lambda = 0.30$,  the adaptive control parameters  $\gamma _1 =5.0$ and $ \gamma_2 = 1.0$, and the LRG parameters  $\varepsilon_0=0.055$ and $k_{\varepsilon} = 5$ were chosen.  With these choices, the closed-loop system specified in Theorem 4.1 was simulated. 
The resulting temperature profile and reference trajectories generated by the closed-loop adaptive system are shown in Figures \ref{fig:temperature_profile} and \ref{fig:temperature_reference_model}.
The results show the superiority of the proposed control design.  
\FloatBarrier
\FloatBarrier

\section{Conclusion}

This paper presents an adaptive control architecture that integrates adaptive control with reference governors. In the proposed approach, an LRG designed for a reference model with time-varying constraints provides a trajectory for the closed-loop adaptive system to follow, ensuring the satisfaction of state constraints. The adaptive controller guarantees stability and maintains performance despite parametric uncertainties.

The controller was applied to temperature regulation in a fuel cell system where there are parametric uncertainties. Validation via numerical simulation using a bilinear fuel cell model demonstrates that the proposed architecture successfully meets both control and safety objectives, whereas a standard adaptive controller only meets the control objective. 

While the current results demonstrate the effectiveness of the proposed architecture, achieving the simultaneous satisfaction of both input and state constraints remains an area of ongoing research. Future work will investigate methods for ensuring feasibility of the underlying optimization problem in LRGS, and in constructing larger admissible sets to reduce the inherent conservatism of Lyapunov-based invariant sets. 

\begin{appendices}
    \section{}

The total power produced by the electrochemical reaction $\dot Q_{gen}$ is given by \eqref{eq: pemfc_total_power} where $\dot m_{H_2}$ is the mass flow rate of hydrogen and $\Delta H$ is the total enthalpy of combustion of hydrogen. The load current $I_{cell}$ is given by \eqref{eq: pemfc_current} where $\eta_I$ is the Faraday efficiency, $N_{cell}$ is the number of cells in the fuel cell stack, $MM_{H_2}$ is the molar mass of hydrogen, and $F$ is the Faraday constant. The electrical power of the fuel cell is given by Ohm's Law in \eqref{eq: pemfc_electrical_power} where $V_{cell}$ is the cell voltage.  

To determine $V_{cell}$, we leverage the empirical model developed in \cite{Ballard}  that expresses the cell voltage as a function of current and stack temperature $T_{st}(t)$. The cell voltage is given in \eqref{eq: pemfc_voltage} where $E_0$ represents the open circuit voltage of the fuel cell, $V_{act}, V_{ohm}, V_{trans}$ represent the overpotentials corresponding to activation, ohmic, and transport losses, respectively. The overpotentials are given by
\begin{equation*}
    V_{act} = \Big(\alpha_1 + \alpha_2T_{st}(t)\Big)\ln{\Big(\frac{I_{cell}}{A_{cell}}\Big)}
\end{equation*}
\begin{equation*}
    V_{ohm} = \Big(\beta_1 +\beta_2T_{st}(t)\Big)\frac{I_{cell}}{A_{cell}}
\end{equation*}
\begin{equation*}
    V_{trans} = \Big(\theta_1 + \theta_2T_{st}(t)\Big)\exp{\Big(n\frac{I_{cell}}{A_{cell}}\Big)}
\end{equation*}
where $A_{cell}$ is the fuel cell surface area, $\alpha_1, \alpha_2, \beta_1, \beta_2, \theta_1, \theta_2$, and $n$ are empirical parameters that characterizes the MK5-PEM Ballard fuel cell studied in \cite{Ballard}. 

The heat removed from the fuel cell by coolant is given in \eqref{eq: pemfc_coolant_power} where $w_c$ is the mass flow of the coolant, $c_{p_c}$ is the heat specific heat capacity of the coolant, and $T_{in}$ is the  temperature of the coolant into the fuel cell stack in the PEMFC system. A radiator is used to remove heat from the coolant coming from fuel cell stack. The temperature at the radiator outlet is assumed to be regulated by another controller and assumed to be at a constant value. We simplify \eqref{eq: pemfc_thermal_balance} as \eqref{eq: pemfc_bilinear} where
\begin{equation*}
    \begin{split}
        A_0(I_{cell}) = \frac{N_{cell}I_{cell}}{m_{st}C_{p_{st}}}\Bigg(\frac{\eta_I MM_{H_2}\Delta H}{2F} + \frac{1}{1000}\Bigg(-E_0 + \alpha_1 \ln \Big(\frac{1000I_{cell}}{A_{cell}}\Big) + \beta_1 \Big(\frac{1000I_{cell}}{A_{cell}}\Big) \\+ \theta_1 \exp \Big(n\frac{1000I_{cell}}{A_{cell}}\Big)\Bigg)\Bigg)
    \end{split}
\end{equation*}
\begin{equation*}
    \begin{split}
          A_1(I_{cell}) = \frac{N_{cell}I_{cell}}{1000m_{st}C_{p_{st}}}\Bigg(\alpha_2\ln\Big(\frac{1000I_{cell}}{A_{cell}}\Big)  + \beta_2 \Big(\frac{1000I_{cell}}{A_{cell}}\Big) + \theta_2 \exp{\Big(n\frac{1000I_{cell}}{A_{cell}}\Big)}\Bigg)
    \end{split}
\end{equation*}
\begin{equation*}
    B_0 = \frac{C_{p_c}}{m_{st}C_{p_{st}}}
\end{equation*}
\begin{table}[htbp]
    \centering
    \begin{tabular}{c c c c}
    \hline
    Parameter & Value & Parameter & Value \\
    \hline 
        $\alpha_1$ [\si{\volt}] & \num{4.01e-2} & $A_{cell}$ [\si{\centi\meter\squared}]  &  \num{232}\\
        $\alpha_2$ [\si{\volt\per\celsius}] & \num{-1.40e-4} & $N_{cell}$ & \num{36}\\ 
        $\beta_1$  [\si{\kilo\ohm\centi\meter\squared}] & \num{4.77e-4} & $m_{st}C_{p_{st}}$ [\si{\kilo\joule\per\celsius}] & \num{35} \\
        $\beta_2$  [\si{\kilo\ohm\centi\meter\squared\per\celsius}] & \num{-3.32e-6} & $F$ [\si{\coulomb \per\mole}] & \num{96485} \\
        $\theta_1$ [\si{\volt}] & \num{1.1e-4} & $MM_{H_2}$ [\si{\gram\per\mole}] & \num{2.016}\\
        $\theta_2$ [\si{\volt\per\celsius}]& \num{-1.2e-6} & $\Delta H$ [\si{\kilo\joule\per\gram}] & \num{143} \\
        $n$  [\si{\centi\meter\squared\per\milli\ampere}] & \num{8.0e-3} & $\eta_1$ & \num{0.98} \\
        $E_{0}$ [\si{\volt}] & \num{1.05} & $C_{p_c}$ [\si{\kilo\joule\per\kilo\gram\per\celsius}] & \num{4.184} \\ 
    \hline
    \end{tabular}
    \caption{Parameters of the fuel cell model}
    \label{tab:Parameters of the fuel cell model}
\end{table}
\label{app: pemfc_model}

    \section{}

\textbf{Lemma 1.} For the linear dynamic system given by \eqref{eq: refmodel}
where (i) $A_m$ is Hurwitz and (ii) $x_d$ is a constant, the steady-state value of $x_m$, $\bar x_m$, is given by 
\begin{equation*}
    \bar x_m = -\begin{bmatrix}
        1 & 0
    \end{bmatrix}A_m^{-1}b_m x_d  =x_d.
\end{equation*}
\textbf{Proof.} We omit the proof as it is straightforward. 
\newline \textbf{Lemma 2.} For the linear dynamic system given by
\begin{equation}\label{eq: lemma2_system}
    \dot z_m = A_m z_m + b_m x_d+ \begin{bmatrix}
        1 \\ 0
    \end{bmatrix}\dot x_d 
\end{equation}
where (i) $A_m$ is Hurwitz, (ii) $ x_d(t) \rightarrow \bar x_d$ as $t\rightarrow \infty$, where $\bar x_d$ is a constant, and (iii) $|\dot x_d(t)| \leq \rho$, the tracking error $e_m$ defined as 
\begin{equation*}
    e_m = \bar x_d - x_m,
\end{equation*}
satisfies the inequality $\lim _{t\rightarrow \infty}|e_m(t)| \leq  k\rho$, where $k$ is the induced norm of the linear system \eqref{eq: lemma2_system}. 
\newline \textbf{Proof.} The proof is omitted as it is straightforward.

    \section{}
\subsection{Proof of Result (i)}
The Lyapunov function candidate in \eqref{eq: lyapunov_function_candidate}, its derivative defined in \eqref{eq: lyapunov_function_derivative}, and the parameter adaptation law \eqref{eq: adaptive_laws} guarantee that $V(t) \leq V(0)$ $\forall t\geq 0$.  This implies that $e_2 (t), \widehat J(t)$, and $\widehat B(t)$ are bounded. Proposition 2 implies that $\tilde x^{'}_d$ is bounded. This property together with the fact that $A_m$ is Hurwitz implies that $x_m(t)$ and $\dot x_m (t)$ are bounded. The boundedness of $e_2(t)$, implies that $e(t)$ is bounded. It follows from the boundedness of $e_2(t)$ and $x_m(t)$ that $x(t)$ is bounded. The boundedness and $e_2(t)$ and $\dot x_m(t)$ ensures that $e_1(t)$ is bounded. Therefore, the states of the closed-loop adaptive system are all bounded. 

\subsection{Proof of Result (ii)}
 The boundedness of $e_2(t), e_1(t)$, and $x(t)$ guarantees that $\dot e_2(t)$ is bounded. This further implies, from Barbalat's lemma that  $\lim _{t\rightarrow\infty} e_2(t) = 0$. Therefore, 
 $x(t)$ tracks $x_m(t)$ asymptotically. From Lemma 2 and from the sufficient condition in Theorem 4.1, 
 the control objective is achieved. 

\subsection{Proof of Result (iii)}
Assuming the the solution to the LRG is feasible, the LRG guarantees $x_m \in \mathcal{C}_{\bar x - \varepsilon}$. Using algebraic simplifications, we have
\begin{equation}
    -\bar x + (\varepsilon_0 + k_{\varepsilon}e^2 - e)\leq x \leq \bar x -(\varepsilon_0 + k_{\varepsilon}e^2+e).
\end{equation}
To prove that $x \in C_{\bar x - \varepsilon^{'}_0}$, we choose  $k_\varepsilon$ , $\varepsilon_0$, and $\delta$ such that
\begin{equation*}
    0 < \frac{1}{4k_\varepsilon} < \delta < \varepsilon_0
\end{equation*}
which establishes result (iii) in Theorem 1. 

\end{appendices}

\printbibliography
\end{document}